\documentclass[twocolumn,
               superscriptaddress,
               floatfix,
               longbibliography, 
               showkeys,apl]{revtex4-1}

\usepackage{graphicx} 
\usepackage{bm}	
\usepackage{color}                     
\usepackage{xcolor}
\usepackage{epsfig}
\usepackage{amsmath} 
\usepackage{amssymb} 
\usepackage{longtable} 
\usepackage{float}
\usepackage{mathtools}
\usepackage{xparse}
\usepackage{hyperref}
\usepackage{times}
\usepackage[normalem]{ulem}
\usepackage[ruled]{algorithm}
\usepackage{algpseudocode}
\usepackage{enumitem}

\usepackage{sidecap}
\sidecaptionvpos{figure}{t}

\usepackage{tabularx}
\usepackage{graphicx}
\usepackage{adjustbox}

\newcommand{\be}{\begin{equation}}
\newcommand{\ee}{\end{equation}}
\newcommand{\bd}{\begin{displaymath}}
\newcommand{\ed}{\end{displaymath}}
\newcommand{\BE}{\begin{eqnarray}}
\newcommand{\EE}{\end{eqnarray}}

\definecolor{darkgreen}{rgb}{0.0, 0.5, 0.0}

\DeclareMathOperator*{\argmin}{arg\,min}
\DeclareMathOperator*{\argmax}{arg\,max}

\begin{document}

\setlist[enumerate,1]{label=\arabic*, start=0}

\title{GEO: Enhancing Combinatorial Optimization with Classical and Quantum Generative Models}

\author{Javier Alcazar}
\affiliation{Zapata Computing Canada Inc., 325 Front St W, Toronto, ON, M5V 2Y1}

\author{Mohammad Ghazi Vakili}
\affiliation{Zapata Computing Canada Inc., 325 Front St W, Toronto, ON, M5V 2Y1}
\affiliation{Department of Chemistry, University of Toronto, Toronto, ON, M5G 1Z8, Canada}
\affiliation{Department of Computer Science, University of Toronto, Toronto, Ontario M5S 2E4, Canada}

\author{Can B. Kalayci}
\affiliation{Zapata Computing Canada Inc., 325 Front St W, Toronto, ON, M5V 2Y1}
\affiliation{Department of Industrial Engineering, Pamukkale University, Kinikli Campus, 20160, Denizli, Turkey}

\author{Alejandro Perdomo-Ortiz}
\email{alejandro@zapatacomputing.com}
\affiliation{Zapata Computing Canada Inc., 325 Front St W, Toronto, ON, M5V 2Y1}


\date{\today} 

\begin{abstract}
We introduce a new framework that leverages machine learning models known as generative models to solve optimization problems. Our Generator-Enhanced Optimization (GEO) strategy is flexible to adopt any generative model, from quantum to quantum-inspired or classical, such as Generative Adversarial Networks, Variational Autoencoders, or Quantum Circuit Born Machines, to name a few. Here, we focus on a quantum-inspired version of GEO relying on tensor-network Born machines, and referred to hereafter as TN-GEO. We present two prominent strategies for using TN-GEO. The first uses data points previously evaluated by any quantum or classical optimizer, and we show how TN-GEO improves the performance of the classical solver as a standalone strategy in hard-to-solve instances. The second strategy uses TN-GEO as a standalone solver, i.e., when no previous observations are available. Here, we show its superior performance when the goal is to find the best minimum given a fixed budget for the number of function calls. This might be ideal in situations where the cost function evaluation can be very expensive. To illustrate our results, we run these benchmarks in the context of the portfolio optimization problem by constructing instances from the S\&P 500 and several other financial stock indexes. We show that TN-GEO can propose unseen candidates with lower cost function values than the candidates seen by classical solvers. This is the first demonstration of the generalization capabilities of quantum-inspired generative models that provide real value in the context of an industrial application. We also comprehensively compare state-of-the-art algorithms in a generalized version of the portfolio optimization problem. The results show that TN-GEO is among the best compared to these state-of-the-art algorithms; a remarkable outcome given the solvers used in the comparison have been fine-tuned for decades in this real-world industrial application. We see this as an important step toward a practical advantage with quantum-inspired models and, subsequently, with quantum generative models.

\end{abstract}

\maketitle

\section{Introduction}\label{s:intro}
Along with machine learning and the simulation of materials, combinatorial optimization is one of top candidates for practical quantum advantage. That is, the moment where a quantum-assisted algorithm outperforms the best classical algorithms in the context of a real-world application with a  commercial or scientific value. There is an ongoing portfolio of techniques to tackle optimization problems with quantum subroutines, ranging from algorithms tailored for quantum annealers (e.g., Refs.~\cite{kadowaki_quantum_1998, Farhi2001}), gate-based quantum computers (e.g., Refs.~\cite{Farhi2014, Hadfield2017}) and quantum-inspired (QI) models based on tensor networks (e.g., Ref.~\cite{mugel2020dynamic}). 

Regardless of the quantum optimization approach proposed to date, there is a need to translate the real-world problem into a polynomial unconstrained binary optimization (PUBO) expression -- a task which is not necessarily straightforward and that usually results in an overhead in terms of the number of variables. Specific real-world use cases illustrating these PUBO mappings are depicted in Refs.~\cite{PerdomoOrtiz2012_LPF} and \cite{PerdomoOrtiz2017a}.  Therefore, to achieve practical quantum advantage in the near-term, it would be ideal to find a quantum optimization strategy that can work on arbitrary objective functions, bypassing the translation and overhead limitations raised here.

In our work, we offer a solution to these challenges by proposing a novel generator-enhanced optimization (GEO) framework which leverage the power of (quantum or classical) generative models. This family of solvers can scale to large problems where combinatorial problems become intractable in real-world settings. Since our optimization strategy does not rely on the details of the objective function to be minimized, it is categorized in the group of so-called \textit{black-box solvers}. Another highlight of our approach is that it can utilize available observations obtained from attempts to solve the optimization problem. These initial evaluations can come from any source, from random search trials to tailored state-of-the-art (SOTA) classical or quantum optimizers for the specific problem at hand.

\begin{figure*}
\includegraphics[width=\textwidth, scale=0.5]{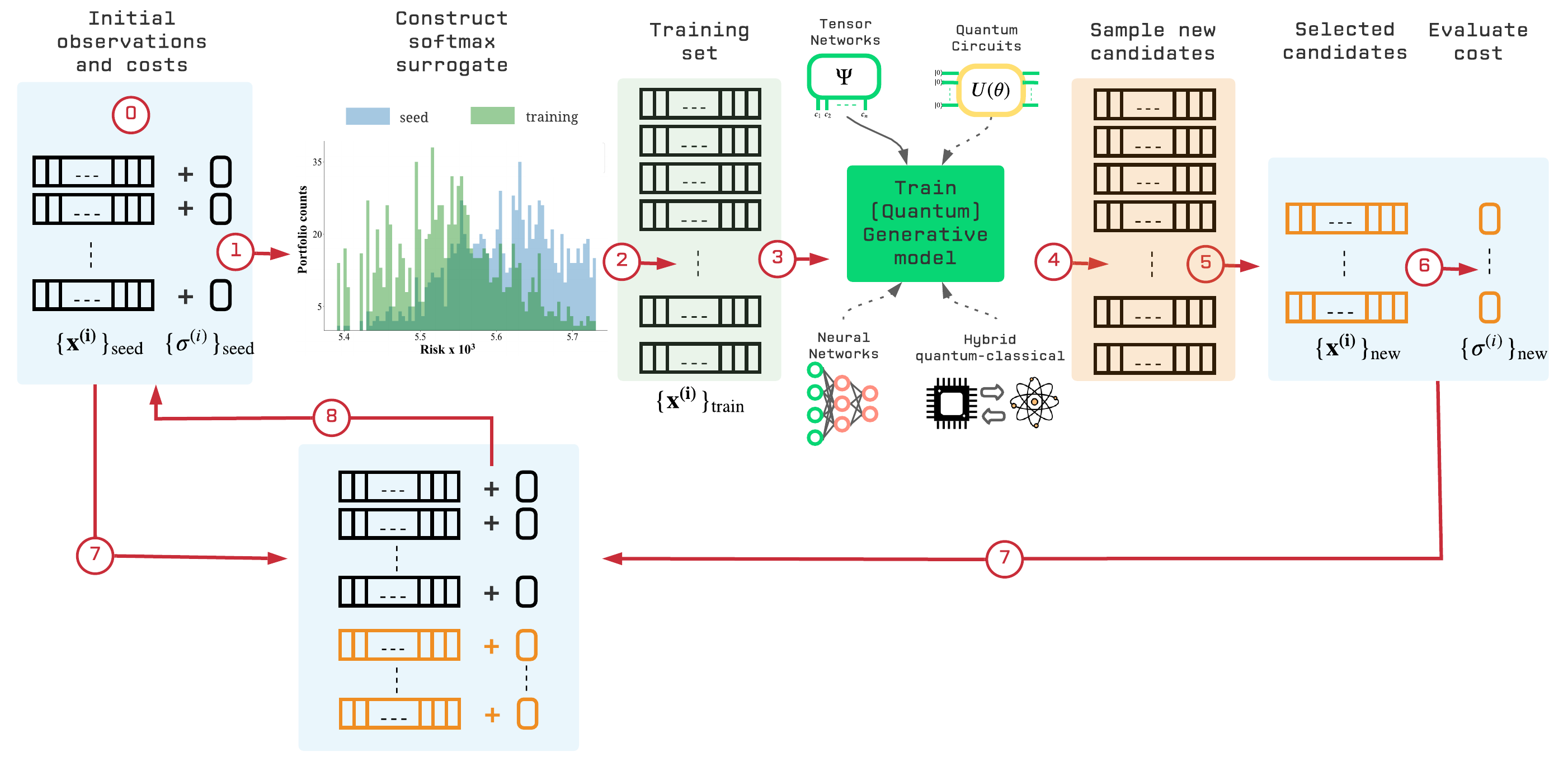}
\caption{\textbf{Scheme for our Generator-Enhanced Optimization (GEO) strategy.} The GEO framework leverages generative models to utilize previous samples coming from any quantum or classical solver. The trained quantum or classical generator is responsible for proposing candidate solutions which might be out of reach for conventional solvers. This \textit{seed data set} (step 0) consists of observation bitstrings $\{\boldsymbol{x}^{(i)}\}_{ \rm{seed} }$ and their respective costs $\{ \sigma^{(i)} \}_{ \rm{seed}} $. To give more weight to samples with low cost, the seed samples and their costs are used to construct a \textit{softmax} function which serves as a \textit{surrogate} to the cost function but in probabilistic domain. This softmax surrogate also serves as a prior distribution from which the \textit{training set} samples are withdrawn to train the generative model (steps 1-3). As shown in the figure between steps 1 and 2, training samples from the softmax surrogate are biased favoring those with low cost value. For the work presented here, we implemented a tensor-network (TN)-based generative model. Therefore, we refer to this quantum-inspired instantiation of GEO as TN-GEO. Other families of generative models from classical, quantum, or hybrid quantum-classical can be explored as expounded in the main text. The quantum-inspired generator corresponds  to a tensor-network Born machine (TNBM) model which is used to capture the main features in the training data, and to propose new solution candidates which are subsequently post selected before their costs  $\{ \sigma^{(i)} \}_{ \rm{new}} $ are evaluated (steps 4-6). The new set is merged with the seed data set (step 7) to form an updated seed data set (step 8) which is to be used in the next iteration of the algorithm. More algorithmic details for the two TN-GEO strategies proposed here, as a \textit{booster} or as a \textit{stand-alone} solver, can be found in the main text and in \ref{ss:TN-GEO_booster} and \ref{ss:TN-GEO_stand-alone} respectively.   
}
\label{fig:Algo_Scheme}
\end{figure*}

Our GEO strategy is based on two key ideas. First, the generative-modeling component aims to capture the correlations from the previously observed data (step 0-3 in Fig.~\ref{fig:Algo_Scheme}). Second, since the focus here is on a minimization task, the (quantum) generative models need to be capable of generating  new “unseen” solution candidates which have the potential to have a lower value for the objective function than those already “seen” and used as the training set (step 4-6 in Fig.~\ref{fig:Algo_Scheme}). This exploration towards unseen and valuable samples is by definition the fundamental concept behind generalization: the most desirable and important feature of any practical ML model. We will elaborate next on each of these components and demonstrate these two properties in the context of the tensor-network-based generative models and its application to a non-deterministic polynomial-time hard (NP-hard) version of the portfolio optimization in finance.

 To the best of our knowledge, this is the first optimization strategy proposed to do an efficient \textit{blackbox exploration} of the objective-function landscape with the help of generative models. Although other proposal leveraging generative models as a subroutine within the optimizer have appeared recently since the publication of our manuscript (e.g., see GFlowNets~\cite{Bengio2021} and the variational neural annealing~\cite{HibatAllah2022} algorithms), our framework is the only capable of both, handling arbitrary cost functions and also with the possibility of swapping the generator for a quantum or quantum-inspired implementation. GEO also has the enhanced feature that the more data is available, the more information can be passed and used to train the (quantum) generator.

In this work, we highlight the different features of GEO by performing a comparison with alternative solvers, such as Bayesian optimizers and generic solvers like simulated annealing. In the case of the specific real-world large-scale application of portfolio optimization, we compare against the SOTA optimizers and show the competitiveness of our approach. These results are presented in Sec.~\ref{s:results}. Next, in Sec.~\ref{s:geo}, we present the GEO approach and its range of applicability.

\section{Quantum-Enhanced Optimization with Generative Models}\label{s:geo}

As shown in Fig.~\ref{fig:Algo_Scheme}, depending on the GEO specifics we can construct an entire family of solvers whose generative modeling core range from classical, QI or quantum circuit (QC) enhanced, or hybrid quantum-classical model. These options can be realized by utilizing, for example, Boltzmann machines~\cite{Cheng2017} or Generative Adversarial Networks (GAN)~\cite{goodfellow2014generative}, Tensor-Network Born Machines (TNBM)~\cite{Cheng2017TensorBorn}, Quantum Circuit Born Machines (QCBM)\cite{Benedetti2019} or Quantum-Circuit Associative Adversarial Networks (QC-AAN)\cite{rudolph2020generation} respectively, to name just a few of the many options for this probabilistic component.

QI algorithms come as an interesting alternative since these allow one to simulate larger scale quantum systems with the help of efficient tensor-network (TN) representations. Depending on the complexity of the TN used to build the quantum generative model, one can simulate from thousands of problem variables to a few tens, the latter being the limit of simulating an universal gate-based quantum computing model. This is, one can control the amount of quantum resources available in the quantum generative model by choosing the QI model.

Therefore, from all quantum generative model options, we chose to use a QI generative model based on TNs to test and scale our GEO strategy to instances with a number of variables commensurate with those found in industrial-scale scenarios. We refer to our solver hereafter as \textit{TN-GEO}. For the training of our TN-GEO models we followed the work of Han et al. \cite{Han2018} where they proposed to use Matrix Product States (MPS) to build the unsupervised generative model. The latter extends the scope from early successes of quantum-inspired models in the context of supervised ML~\cite{stoudenmire2016supervised,Stavros2019, Roberts2019,fishman2020itensor}.

In this paper we will discuss two modes of operation for our family of quantum-enhanced solvers: 

\begin{itemize}
    \item  In \textbf{TN-GEO \textit{as a "booster"}} we leverage past observations from classical (or quantum) solvers. To illustrate this mode we use observations from simulated annealing (SA) runs. Simulation details are provided in Appendix~\ref{ss:TN-GEO_booster}.
    \item In \textbf{TN-GEO \textit{as a stand-alone solver}} all initial cost function evaluations are decided entirely by the quantum-inspired generative model, and a random prior is constructed just to give support to the target probability distribution the MPS model is aiming to capture. Simulation details are provided in Appendix~\ref{ss:TN-GEO_stand-alone}. 
\end{itemize}

Both of these strategies are captured in the algorithm workflow diagram in Fig.~\ref{fig:Algo_Scheme} and described in more detail in Appendix~\ref{s:methods}. 

\section{Results and Discussion}~\label{s:results}
To illustrate the implementation for both of these settings we tested their performance on an NP-hard version of the portfolio optimization problem with cardinality constraints.  The selection of optimal investment on a specific set of assets, or \textit{portfolios}, is a problem of great interest in the area of quantitative finance. This problem is of practical importance for investors, whose objective is to allocate capital optimally among assets while respecting some investment restrictions. The goal of this optimization task, introduced by Markowitz~\cite{Markowitz52}, is to generate a set of portfolios that offers either the highest expected return (profit) for a defined level of risk or the lowest risk for a given level of expected return. In this work, we focus in two variants of this cardinality constrained optimization problem. The first scenario aims to choose portfolios which minimize the  volatility or risk given a specific target return (more details are provided in Appendix~\ref{ss:portfolio}.) To compare with the reported results from the best performing SOTA algorithms, we ran TN-GEO in a second scenario where the goal is to choose the best portfolio given a fixed level of \textit{risk aversion}. This is the most commonly used version of this optimization problem when it comes to comparison among SOTA solvers in the literature (more details are provided in Appendix~\ref{apx:comparisonform}).

\subsection{TN-GEO as a booster for any other combinatorial optimization solver}\label{subsectionTNQEO}

\begin{figure}[h!]
\includegraphics[width=\linewidth, scale=0.5]{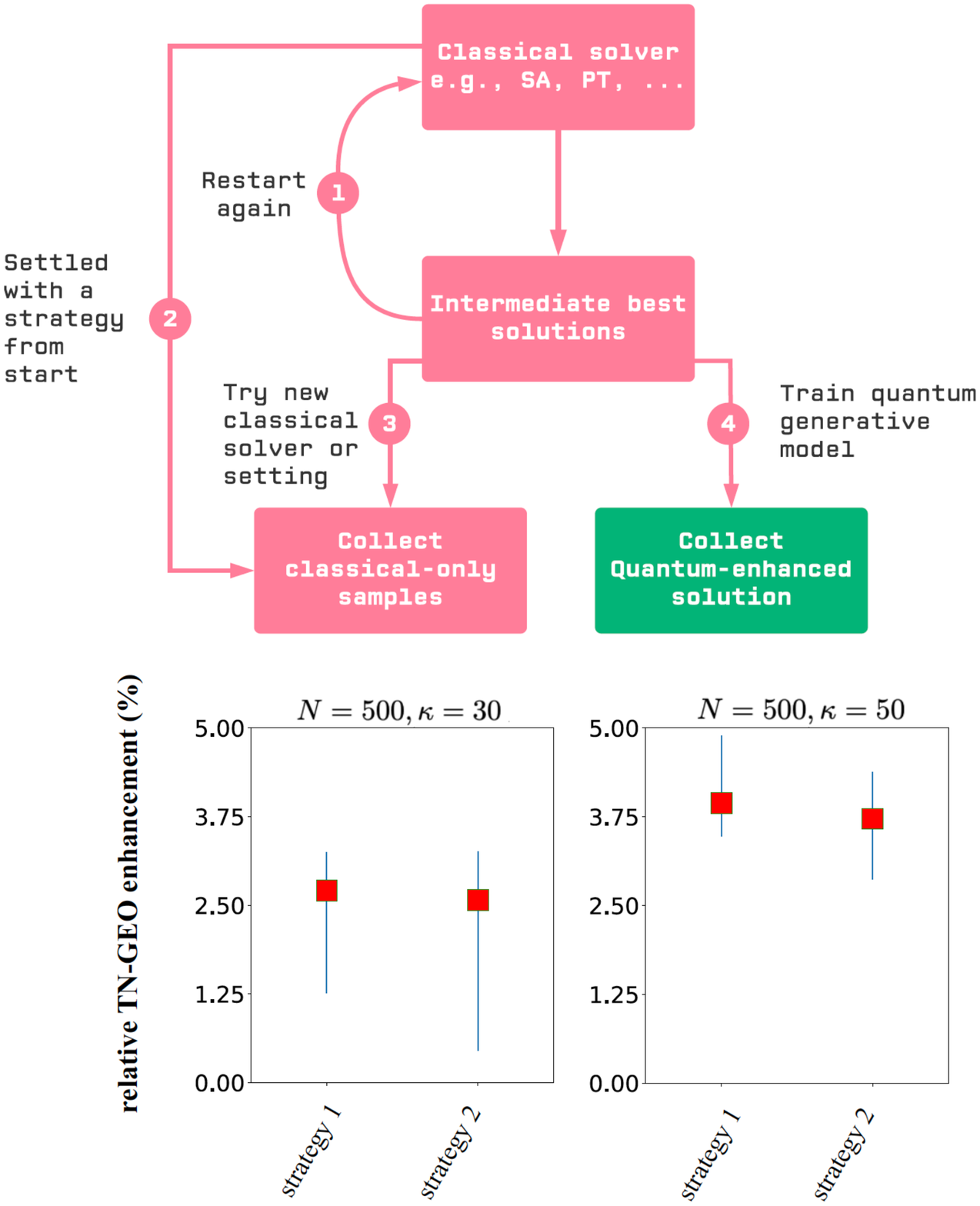}
\caption{\textbf{TN-GEO as a \textit{booster}.} Top: Strategies 1-3 correspond to the current options a user might explore when solving a combinatorial optimization problem with a suite of classical optimizers such as simulated annealing (SA), parallel tempering (PT), generic algorithms (GA), among others. In strategy 1, the user would use its computational budget with a preferred solver. In strategy 2-4 the user would inspect intermediate results and decide whether to keep trying with the same solver (strategy 2), try a new solver or a new setting of the same solver used to obtain the intermediate results (strategy 3), or, as proposed here, to use the acquired data to train a quantum or quantum-inspired generative model within a GEO framework such as TN-GEO (strategy 4). Bottom: Results showing the relative TN-GEO enhancement from TN-GEO  over either strategy 1 or strategy 2. Positive values indicate runs where TN-GEO outperformed the respective classical strategies (see Eq.~\ref{eq:rel-qe}). The data represents bootstrapped medians from 20 independent runs of the experiments and error bars correspond to the 95\% confidence intervals. The two instances presented here correspond to portfolio optimization instances where all the assets in the S\&P 500 market index where included ($N=500$), under two different cardinality constraints $\kappa$. This cardinality constraint indicate the number of assets that can be included at a time in valid portfolios, yielding a search space of $M = \binom{N}{\kappa}$, with $M \sim 10^{69}$ portfolios candidates for $\kappa=50$.
}
\label{fig:TN-GEObooster}
\end{figure}

In Fig.~\ref{fig:TN-GEObooster} we present the experimental design and the results obtained from using TN-GEO as a booster. In these experiments we illustrate how using intermediate results from simulated annealing (SA) can be used as seed data for our TN-GEO algorithm. As described in Fig.~\ref{fig:TN-GEObooster}, there are two strategies we explored (strategies 1 and 2) to compare with our TN-GEO strategy (strategy 4). To fairly compare each strategy, we provide each with approximately the same computational wall-clock time. For strategy 2, this translates into performing additional restarts of SA with the time allotted for TN-GEO. In the case of strategy 1, where we explored different settings for SA from the start compared to those used in strategy 2, this amounts to using the same total number of number of cost functions evaluations as those allocated to SA in strategy 2. For our experiments this number was set to 20,000 cost function evaluations for strategies 1 and 2. In strategy 4, the TN-GEO was initialized with a prior consisting of the best 1,000 observations out of the first 10,000 coming from strategy 2 (see Appendix \ref{ss:TN-GEO_booster} for details). To evaluate the performance enhancement obtained from the TN-GEO strategy we compute the \textit{relative TN-GEO enhancement} $\eta$, which we define as 
\begin{equation}\label{eq:rel-qe}
    \eta  = \frac{C^{\rm{cl}}_{ \rm{min} } - C^{\rm{TN-GEO}}_{ \rm{min}}  }{ C^{ \rm{cl} }_{ \rm{min}} } \times 100\%.
\end{equation}

Here, $C^{ \rm{cl} }_{ \rm{min}}$ is the lowest minimum value found by the classical strategy (e.g., strategies 1-3) while $C^{ \rm{TN-GEO} }_{ \rm{min}}$ corresponds to the lowest value found with the quantum-enhanced approach (e.g., with TN-GEO). Therefore, positive values reflect an improvement over the classical-only approaches, while negative values indicate cases where the classical solvers outperform the quantum-enhanced proposal. 

\begin{figure*}
\includegraphics[width=\textwidth, scale=0.6]{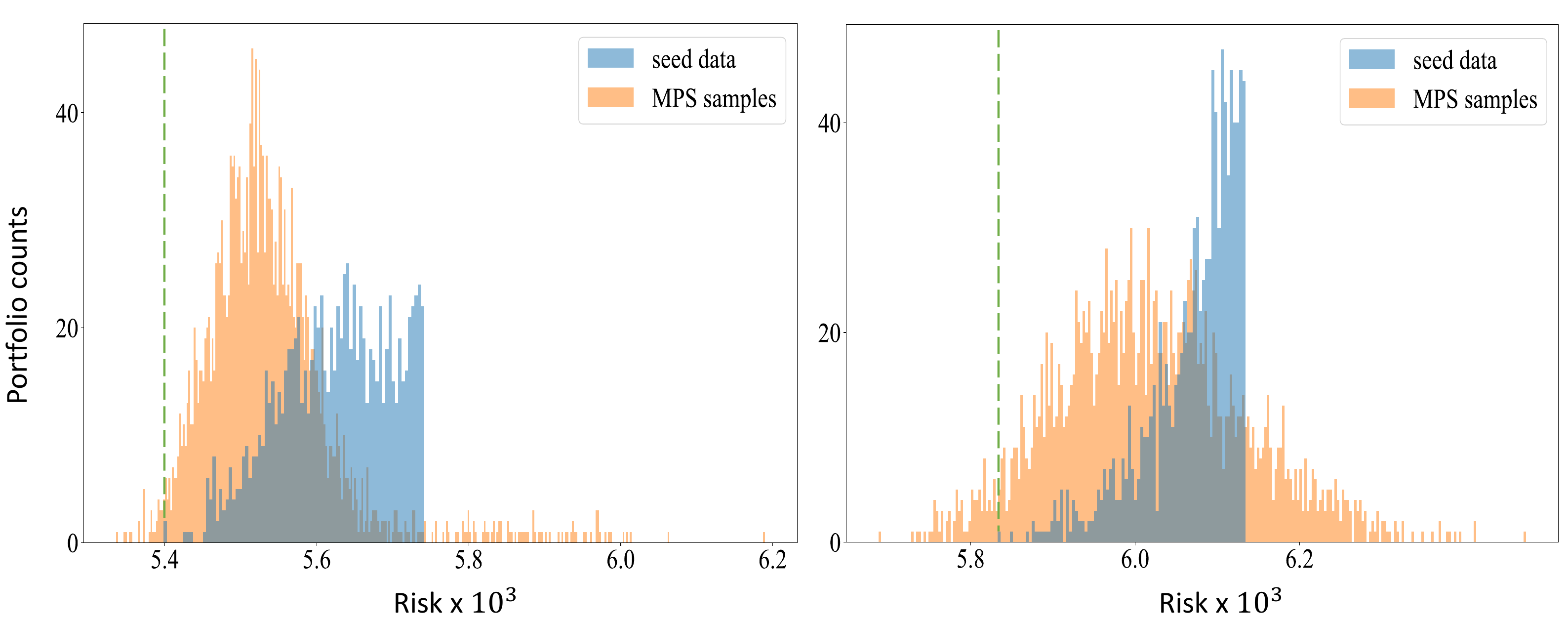}
\caption{\textbf{Generalization capabilities of our quantum-inspired generative model}. Left panel corresponds to an investment universe with $N=50$ assets while the right panel corresponds to one with $N=100$ assets. The blue histogram represents the number of observations or portfolios obtained from the classical solver (\textit{seed data set}). In orange we represent samples coming from our quantum generative model at the core of TN-GEO. The green dash line is positioned at the best risk value found in the seed data. This mark emphasizes all the new outstanding samples obtained with the quantum generative model and which correspond to lower portfolio risk value (better minima) than those available from the classical solver by itself. The number of outstanding samples in the case of $N=50$ is equal to 31, while 349 outstanding samples were obtained from the MPS generative model in the case of $N=100$.  }
\label{fig:Densities_MPS_seed2}
\end{figure*}

As shown in the  Fig.~\ref{fig:TN-GEObooster}, we observe that  TN-GEO outperforms on average both of the classical-only strategies implemented. The quantum-inspired enhancement observed here, as well as the trend for a larger enhancement as the number of variables (assets) becomes larger, is confirmed in many other investment universes with a number of variables ranging from $N=30$ to $N=100$  (see Appendix~\ref{apx:relQE} for more details).  Although we show an enhancement compared to SA, similar results could be expected when other solvers are used, since our approach builds on solutions found by the solver and does not compete with it from the start of the search. Furthermore, the more data available, the better the expected performance of TN-GEO is. An important highlight of TN-GEO as a booster is that these previous observations can come from a combination of solvers, as different as purely quantum or classical, or hybrid. 

The observed performance enhancement compared with the classical-only strategy must be coming from a better exploration of the relevant search space, i.e., the space of those bitstring configurations $\boldsymbol{x}$ representing portfolios which could yield a low risk value for a specified expected investment return. That is the intuition behind the construction of TN-GEO. The goal of the generative model is to capture the important correlations in the previously observed data, and to use its generative capabilities to propose similar new candidates.

Generating new candidates is by no means a trivial task in ML and it determines the usefulness and power of the model since it measure its \textit{generalization} capabilities.  In this setting of QI generative models, one expects that the MPS-based generative model at the core of TN-GEO is not simply memorizing the observations given as part of the training set, but that it will provide new unseen candidates. This is an idea which has been recently tested and demonstrated to some extent on synthetic data sets (see e.g., Refs.~\cite{Bradley_2020}, \cite{e21121236} and \cite{miller2020tensor}. In Fig.~\ref{fig:Densities_MPS_seed2} we demonstrate that our quantum-inspired generative model is generalizing to new samples and that these add real value to the optimization search. To the best of our knowledge this is the first demonstration of the generalization capabilities of quantum generative models in the context of a real-world application in an industrial scale setting, and one of our main findings in our paper.

Note that our TN-based generative model not only produces better minima than the classical seed data, but it also generates a rich amount of samples in the low cost spectrum. This bias is imprinted in the design of our TN-GEO and it is the purpose of the \textit{softmax} surrogate prior distribution shown in Fig.~\ref{fig:Algo_Scheme}. This richness of new samples could be useful not only for the next iteration of the algorithm, but they may also be readily of value to the user solving the application. In some applications there is value as well in having information about the runners-up. Ultimately, the cost function is just a model of the system guiding the search, and the lowest cost does not translate to the best performance in the real-life investment strategy.

\subsection{Generator-Enhanced Optimization as a Stand-Alone Solver}

\begin{figure*}
\includegraphics[width= 0.95
\textwidth,scale=0.7]{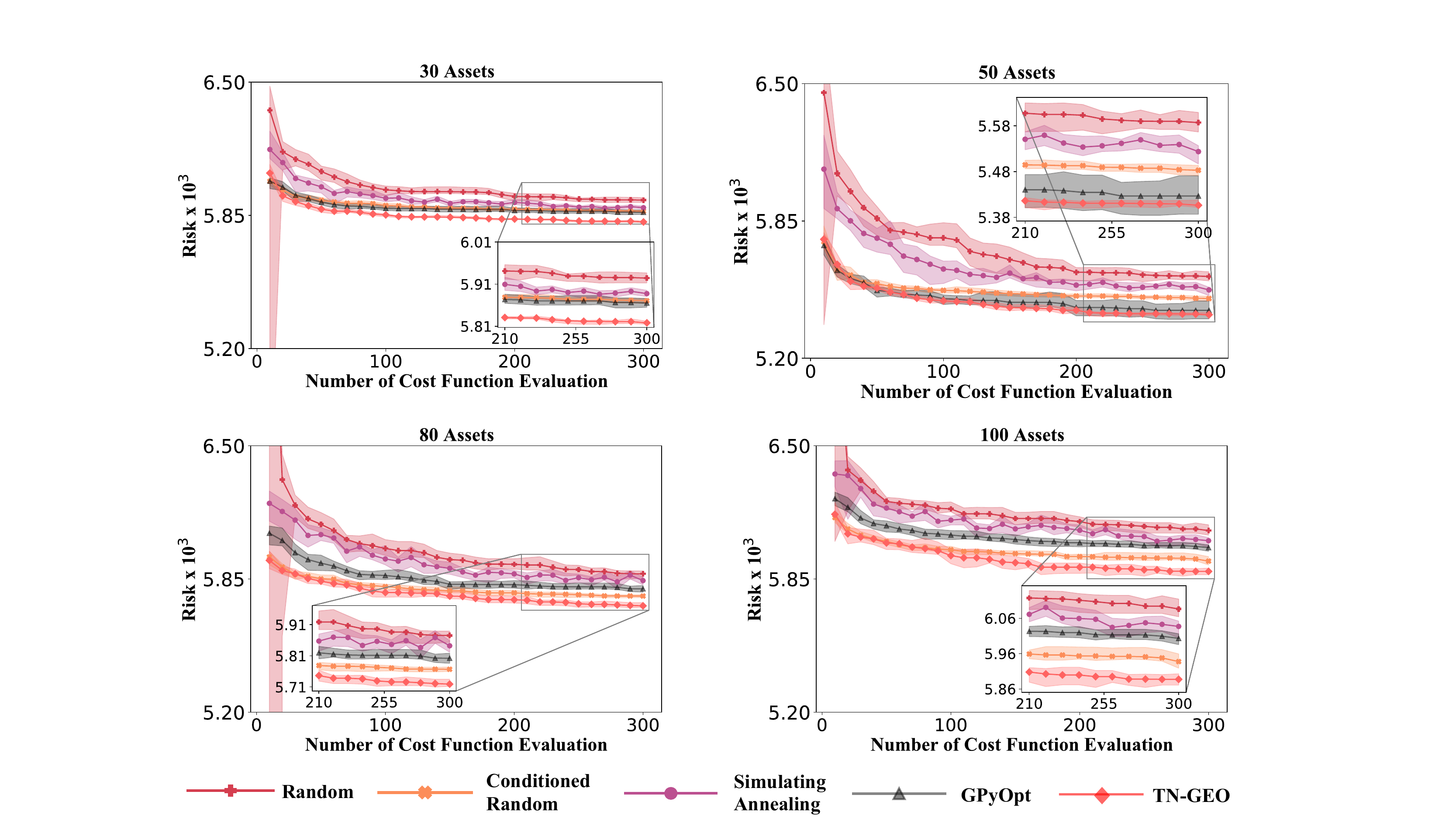}
\caption{\textbf{TN-GEO as a \textit{stand-alone} solver:} In this comparison of TN-GEO against four classical competing strategies, investment universes are constructed from subsets of the S\&P 500 with a diversity in the number of assets (problem variables) ranging from $N=30$ to $N=100$. The goal is to minimize the risk given an expected return which is one of the specifications in the combinatorial problem addressed here. Error bars and their 95\% confidence intervals are calculated from bootstrapping over 100 independent random initializations for each solver on each problem. The main line for each solver corresponds to the bootstrapped median over these 100 repetitions, demonstrating the superior performance of TN-GEO over the classical solvers considered here. As specified in the text, with the exception of TN-GEO, the classical solvers use to their advantage the \textit{a priori} information coming from the cardinality constraint imposed in the selection of valid portfolios.}
\label{fig:Results_TN-GEO2_Seed2}
\end{figure*}

Next, we explore the performance of our TN-GEO framework as a stand-alone solver. The focus is in combinatorial problems whose cost functions are expensive to evaluate and where finding the best minimum within the least number of calls to this function is desired. In Fig.~\ref{fig:Results_TN-GEO2_Seed2} we present the comparison against four different classical optimization strategies. As the first solver, we use the \textit{random} solver, which corresponds to a fully random search strategy  over the $2^N$ bitstrings of all possible portfolios, where $N$ is the number of assets in our investment universe. As second solver, we use the \textit{conditioned random} solver, which is a more sophisticated random strategy compared to the fully random search. The conditioned random strategy uses the \textit{a priori} information that the search is restricted to bitstrings containing a fixed number of $\kappa$ assets. Therefore the number of combinatorial possibilities is $M = \binom{N}{\kappa}$, which is significantly less than $2^N$. As expected, when this information is not used the performance of the random solver over the entire $2^N$ search space is worse. The other two competing strategies considered here are SA and the Bayesian optimization library GPyOpt~\cite{gpyopt2016}. In both of these classical solvers, we adapted their search strategy to impose this cardinality constraint with fixed $\kappa$ as well (details in Appendix.~\ref{ss:optimizers}).  This raises the bar even higher for TN-GEO which is not using that \textit{a priori} information to boost its performance \footnote{Specific adaptions of the MPS generative model could be implemented such that it conserves the number of assets by construction, borrowing ideas from condensed matter physics where one can impose MPS a conservation in the number of particles in the quantum state.}. As explained in Appendix~\ref{ss:TN-GEO_stand-alone}, we only use this information indirectly during the construction of the artificial seed data set which initializes the algorithm (step 0, Fig.~\ref{fig:Algo_Scheme}) , but it is not a strong constraint during the construction of the QI generative model (step 3, Fig.~\ref{fig:Algo_Scheme}) or imposed to generate the new candidate samples coming from it (step 4, Fig.~\ref{fig:Algo_Scheme}). Post selection can be applied \textit{a posteriori} such that only samples with the right cardinality  are considered as valid candidates towards the selected set (step 5, Fig.~\ref{fig:Algo_Scheme}).

In Fig.~\ref{fig:Results_TN-GEO2_Seed2} we demonstrate the advantage of our TN-GEO stand-alone strategy compared to any of these widely-used solvers. In particular, it is interesting to note that the gap between TN-GEO and the other solvers seems to be larger for larger number of variables.  

\subsection{Comparison with state-of-the-art algorithms}
Finally, we compare TN-GEO with nine different leading SOTA optimizers covering a broad spectrum of algorithmic strategies for this specific combinatorial problem, based on and referred hereafter as: 1) GTS \cite{chang2000heuristics}, the genetic algorithms, tabu search, and simulated annealing; 2) IPSO \cite{deng2012markowitz}, an improved particle swarm optimization algorithm~\cite{deng2012markowitz}; 3) IPSO-SA \cite{mozafari2011new}, a hybrid algorithm combining particle swarm optimization and simulated annealing; 4) PBILD \cite{lwin2013hybrid}, a population-based incremental learning and differential evolution algorithm; 5) GRASP \cite{baykasouglu2015grasp}, a greedy randomized adaptive solution procedure; 6) ABCFEIT \cite{kalayci2017artificial}, an artificial bee colony algorithm with feasibility enforcement and infeasibility toleration procedures; 7)  HAAG \cite{kalayci2020efficient}, a hybrid algorithm integrating ant colony optimization, artificial bee colony and genetic algorithms; 8) VNSQP \cite{akbay2020parallel}, a variable neighborhood search algorithm combined with quadratic programming;  and, 9) RCABC \cite{cura2021rapidly}, a rapidly converging artificial bee colony algorithm.

The test data used by the vast majority of researchers in the literature who have addressed the problem of cardinality-constrained portfolio optimization come from OR-Library~\cite{beasley1990or}, which correspond to the weekly prices between March 1992 and September 1997 of the following indexes: Hang Seng in Hong Kong (31 assets); DAX 100 in Germany (85 assets); FTSE 100 in the United Kingdom (89 assets); S\&P 100 in the United States (98 assets); and Nikkei 225 in Japan (225 assets).

Here we present the results obtained with TN-GEO and its comparison with the nine different SOTA metaheuristic algorithms mentioned above and whose results are publicly available from the literature. Table \ref{table:1} shows the results of all algorithms and  all performance metrics for each of the 5 index data sets (for more details on the evaluation metrics, see  Appendix~\ref{apx:comparisonform}). Each algorithm corresponds to a different column, with TN-GEO in the rightmost column. The values are shown in \textcolor{red}{red} if the TN-GEO algorithm performed better or equally well compared to the other algorithms on the corresponding performance metric. The numbers in \textbf{bold} mean that the algorithm found the best (lowest) value across all algorithms.

From all the entries in this table, 67\% of them correspond to red entries, where TN-GEO either wins or draws, which is a significant percentage giving that these optimizers are among the best reported in the last decades. 


In Table \ref{table:2}  we show a pairwise comparison of TN-GEO against each of the SOTA optimizers. This table reports the number of times TN-GEO wins, loses, or draws compared to results reported for the other optimizer, across all the performance metrics and for all the 5 different market indexes. Note that since not all the performance metrics are reported for all the solvers and market indexes, the total number of wins, draws, or losses varies. Therefore, we report in the same table the overall percentage of wins plus draws in each case. We see that this percentage is greater than 50\% in all the cases.

Furthermore, in Table~\ref{table:2}, we use the Wilcoxon signed-rank test ~\cite{wilcoxon1992individual}, which is a widely used nonparametric statistical test used to evaluate and compare the performance of different algorithms in different benchmarks ~\cite{JMLR:v7:demsar06a}. Therefore, to statistically validate the results, a Wilcoxon signed-rank test is performed to provide a meaningful comparison between the results from TN-GEO algorithm and the SOTA  metaheuristic algorithms. The Wilcoxon signed-rank test tests the null hypothesis that the median of the differences between the results of the algorithms is equal to 0. Thus, it tests whether there is no significant difference between the performance of the algorithms. The null hypothesis is rejected if the significance value ($p$) is less than the significance level ($\alpha$), which means that one of the algorithms performs better than the other. Otherwise, the hypothesis is retained.

As can be seen from the table, the TN-GEO algorithm significantly outperforms the GTS and PBILD methods on all performance metrics rejecting the null hypothesis at the $0.05$ significance level. On the other hand, the null hypotheses are accepted at $\alpha = 0.05$ for the TN-GEO algorithm over the other remaining algorithms. Thus, in terms of performance on all metrics combined, the results show that there is no significant difference between  TN-GEO and these remaining seven SOTA optimizers (IPSO, IPSO-SA, GRASP, ABCFEIT, HAAG, VNSQP, and RCABC) 

Overall, the results confirm the competitiveness of our quantum-inspired proposed approach against SOTA metaheuristic algorithms. This is remarkable given that these metaheuristics have been explored and fine-tuned for decades.

\begin{table*}
\caption{Detailed comparison with SOTA algorithms for each of the five index data sets and on seven different performance indicators described in Appendix~\ref{apx:comparisonform}. Entries in \textcolor{red}{red} correspond to cases where TN-GEO performed better or tied compared to the other algorithm. Entries in \textbf{bold}, corresponding to the best (lowest) value, for each specific indicator.}

\centering
\begin{tabular}{||c l c c c c c c c c c c||} 
 \hline
 Data Set & Performance Indicator & GTS & IPSO & IPSO-SA & PBILD & GRASP & ABCFEIT & HAAG & VNSQP & RCABC & TN-GEO \\ [0.5ex] 
 \hline\hline
 Hang Seng & Mean & 1.0957 & 1.0953 & - & \textcolor{red} {1.1431} & \textcolor{red} {1.0965} & 1.0953 & \textcolor{red} {1.0965} & \textcolor{red} {1.0964} & \bf 1.0873 & 1.0958 \\ 
 & Median & \textcolor{red} {1.2181} & - & - & \textcolor{red} {1.2390} & 1.2155 & \textcolor{red} {1.2181} &  \textcolor{red} {1.2181} & 1.2155 & \bf 1.2154 & 1.2181 \\
 & Min & - & - & - & - & \bf \textcolor{red} {0.0000}	& \bf \textcolor{red} {0.0000} & \bf \textcolor{red} {0.0000} & \bf \textcolor{red} {0.0000} & \bf \textcolor{red} {0.0000}	& \bf 0.0000 \\
 & Max & - & - & - & - & \bf \textcolor{red} {1.5538} & \bf \textcolor{red} {1.5538} & \bf \textcolor{red} {1.5538} & \bf \textcolor{red} {1.5538} & \bf \textcolor{red} {1.5538}	& \bf 1.5538 \\
 & MEUCD & - & - & \bf \textcolor{red} {0.0001} & - & \bf \textcolor{red} {0.0001} & \bf \textcolor{red} {0.0001} & \bf \textcolor{red} {0.0001} & \bf \textcolor{red} {0.0001} & \bf \textcolor{red} {0.0001} & \bf 0.0001 \\ 
 & VRE & - & - & 1.6368	& - & \textcolor{red} {1.6400} & \textcolor{red} {1.6432} & \textcolor{red} {1.6395} & \textcolor{red} {1.6397} & \bf 1.6342 & 1.6392 \\
 & MRE & - & - & 0.6059 & -	& 0.6060 & 0.6047 & \textcolor{red} {0.6085} & 0.6058 & \bf 0.5964 & 0.6082 \\
 \hline
DAX100 & Mean & \textcolor{red} {2.5424}	& \textcolor{red} {2.5417} & - & \textcolor{red} {2.4251} & 2.3126 & \textcolor{red} {2.3258} & 2.3130 & 2.3125 & \bf 2.2898 & 2.3142 \\
 & Median & \bf 2.5466 & - & - & \textcolor{red} {2.5866} & 2.5630 & \textcolor{red} {2.5678} & 2.5587 & 2.5630 & 2.5629 & 2.5660 \\
 & Minimum & - & - & - & - & \textcolor{red} {0.0059} & \textcolor{red} {0.0023} & \textcolor{red} {0.0023} & \textcolor{red} {0.0059} & \textcolor{red} {0.0059}	& \bf 0.0023 \\
 & Maximum & - & - & - & - & \bf \textcolor{red} {4.0275} & \bf \textcolor{red} {4.0275} & \bf \textcolor{red} {4.0275} & \bf \textcolor{red} {4.0275} & \bf \textcolor{red} {4.0275}	& \bf 4.0275 \\
 & MEUCD & - & - &  \bf \textcolor{red} {0.0001} & - & \bf \textcolor{red} {0.0001} & \bf \textcolor{red} {0.0001} & \bf \textcolor{red} {0.0001} & \bf \textcolor{red} {0.0001} & \bf \textcolor{red} {0.0001} & \bf 0.0001 \\
 & VRE & - & - & \textcolor{red} {6.7806}	& - & \textcolor{red} {6.7593} & \textcolor{red} {6.7925} & \textcolor{red} {6.7806} & \textcolor{red} {6.7583} & \textcolor{red} {6.8326} & \bf 6.7540 \\
 & MRE & - & - & \textcolor{red} {1.2770} & -	& \textcolor{red} {1.2769} & 1.2761 & \textcolor{red} {1.2780} & \textcolor{red} {1.2767} & \bf 1.2357 & 1.2763 \\
\hline
FTSE100 & Mean & \textcolor{red} {1.1076} & \textcolor{red} {1.0628} & - & \textcolor{red} {0.9706} & \textcolor{red} {0.8451} & \textcolor{red} {0.8481} & \textcolor{red} {0.8451} & \textcolor{red} {0.8453} & \bf 0.8406 & 0.8445 \\
 & Median & \bf \textcolor{red} {1.0841} & - & - & \bf \textcolor{red} {1.0841} & \bf \textcolor{red} {1.0841} & \bf \textcolor{red} {1.0841} & \bf \textcolor{red} {1.0841} & \bf \textcolor{red} {1.0841} & \bf \textcolor{red} {1.0841} & \bf 1.0841 \\
 & Minimum & - & - & - & - & 0.0016	& \textcolor{red} {0.0047} & \bf 0.0006 & 0.0045 & 0.0016 & 0.0047 \\
 & Maximum & - & - & - & - & \bf 2.0576 & 2.0638 & 2.0605 & 2.0669 & 2.0670 & 2.0775 \\
 & MEUCD & - & - & \bf \textcolor{red} {0.0000} & - & \bf \textcolor{red} {0.0000} & \bf \textcolor{red} {0.0000} & \bf \textcolor{red} {0.0000} & \bf \textcolor{red} {0.0000} & \bf \textcolor{red} {0.0000} & \bf 0.0000 \\
 & VRE & - & - & \textcolor{red} {2.4701} & - & 2.4350 & \textcolor{red} {2.4397} & 2.4350 & 2.4349 & \bf 2.4149 & 2.4342 \\
 & MRE & - & - & 0.3247 & - & 0.3245 & \textcolor{red} {0.3255} & \bf 0.3186 & 0.3252 & 0.3207 & 0.3254 \\
\hline
S\&P100 & Mean & \textcolor{red} {1.9328} & \textcolor{red} {1.6890} & - & \textcolor{red} {1.6386} & \textcolor{red} {1.2937} & \textcolor{red} {1.2930} & \textcolor{red} {1.2930} & \bf 1.2649 & \textcolor{red} {1.3464} & 1.2918 \\
 & Median & \textcolor{red} {1.1823} & - & - & \textcolor{red} {1.1692} & 1.1420 & 1.1369 & \bf 1.1323 & \bf 1.1323 & \textcolor{red} {1.1515} & 1.1452 \\
 & Minimum & - & - & - & - & \textcolor{red} {0.0009} & \bf \textcolor{red} {0.0000} & \bf \textcolor{red} {0.0000} & \bf \textcolor{red} {0.0000} & \textcolor{red} {0.0009} & \bf 0.0000 \\
 & Maximum & - & - & - & - & \textcolor{red} {5.4551} & \bf \textcolor{red} {5.4422} & \textcolor{red} {5.4642} & \textcolor{red} {5.4551} & \textcolor{red} {5.4520} & 5.4422 \\
 & MEUCD & - & - & \bf \textcolor{red} {0.0001} & - & \bf \textcolor{red} {0.0001} & \bf \textcolor{red} {0.0001} & \bf \textcolor{red} {0.0001} & \bf \textcolor{red} {0.0001} & \bf \textcolor{red} {0.0001} & \bf 0.0001 \\
 & VRE & - & - & \textcolor{red} {2.6281} & -	& 2.5211 & 2.5260 & 2.5255 & \bf 2.5105 & \textcolor{red} {2.5364} & 2.5269 \\
 & MRE & - & - & 0.7846 & - & 0.9063 & 0.8885 & \bf 0.7044 & 0.9072 & 0.8858 & 0.9117 \\
\hline
Nikkei & Mean & \textcolor{red} {0.6066} & \textcolor{red} {0.6870} & -	& \textcolor{red} {0.5972} & 0.5782 & 0.5781 & 0.5781 & \textcolor{red} {0.5904} & \bf 0.5665 & 0.5793 \\
 & Median & \textcolor{red} {0.6093} & - & - & \textcolor{red} {0.5896} & \textcolor{red} {0.5857} & \textcolor{red} {0.5856} & 0.5854 & \textcolor{red} {0.5857} & \textcolor{red} {0.5858} & 0.5855 \\
 & Minimum & - & - & - & - & \bf \textcolor{red} {0.0000} & \bf \textcolor{red} {0.0000} & \bf \textcolor{red} {0.0000} & \bf \textcolor{red} {0.0000} & \bf \textcolor{red} {0.0000} & \bf 0.0000 \\
 & Maximum & - & - & - & - & \bf \textcolor{red} {1.1606} & \bf \textcolor{red} {1.1606} & \textcolor{red} {1.1607} & \bf \textcolor{red} {1.1606} & \bf \textcolor{red} {1.1606} & \bf 1.1606 \\
 & MEUCD & - & - & \textcolor{red} {0.0000} & - & \textcolor{red} {0.0000} & \textcolor{red} {0.0000} & \textcolor{red} {0.0000} & \textcolor{red} {0.0000} & \textcolor{red} {0.0000} & 0.0000 \\
 & VRE & - & - & \textcolor{red} {0.9583} & - & \textcolor{red} {0.8359} & \textcolor{red} {0.8396} & \bf 0.8191 & \textcolor{red} {0.8561} & 0.8314 & 0.8353 \\
 & MRE & - & - & \textcolor{red} {1.7090} & - & 0.4184 & 0.4147 & \textcolor{red} {0.4233} & 0.4217 & \bf 0.4042 & 0.4229\\
\hline
\end{tabular}
\label{table:1}
\end{table*}

\begin{table*}
\caption{Pairwise comparison of TN-GEO against each of the SOTA optimizers. The asymptotic significance is part of the Wilcoxon signed-rank test results.
The null hypothesis that the performance of the two algorithms is the same is tested at the $95\%$ confidence level (significance level: $\alpha = .05$). Results show that TN-GEO is on par with all the SOTA algorithms, and in two cases, GTS and PBILD, it significantly outperforms them. We also report the count for TN-GEO wins, losses, and ties, compared to each of the other algorithms.}
\centering
\begin{adjustbox}{width=1\textwidth}{
\begin{tabular}{||l c c c c c c c c c||} 
 \hline
 TN-GEO vs Other: & GTS & IPSO & IPSO-SA & PBILD & GRASP & ABCFEIT & HAAG & VNSQP & RCABC \\
 \hline\hline
  Wins(+) & 6 & 4 & 6 & 9 & 12 & 10 & 11 & 11 & 8\\
  Loss(-) & 2 & 1 & 4 & 0 & 12 & 9 & 11 & 12 & 16\\
  Ties & 2 & 0 & 5 & 1 & 11 & 16 & 13 & 12 & 11\\
  (Wins+Ties)/Total & 80\% & 80\%	& 67\% & 100\% & 66\% & 74\% & 69\%	& 66\% & 54\% \\
  Asymptotic significance ($p$) & .036 & .080 & .308 & .008 & .247 & .888 & .363 & .594 & .110 \\
  Decision & Reject & Retain & Retain & Reject & Retain & Retain & Retain & Retain & Retain  \\
 \hline
\end{tabular}}
\end{adjustbox}

\label{table:2}
\end{table*}


\section{Outlook}

Compared to other quantum optimization strategies, an important feature of TN-GEO is its algorithmic flexibility. As shown here, unlike other proposals, our GEO framework can be applied to arbitrary cost functions, which opens the possibility of new applications that cannot be easily addressed by an explicit mapping to a polynomial unconstrained binary optimization (PUBO) problem. Our approach is also flexible with respect to the source of the seed samples, as they can come from any solver, possibly more efficient or even application-specific optimizers. The demonstrated generalization capabilities of the generative model that forms its core, helps TN-GEO build on the progress of previous experiments with other state-of-the-art solvers, and it provides new candidates that the classical optimizer may not be able to achieve on its own. We are optimistic that this flexible approach will open up the broad applicability of quantum and quantum-inspired generative models to real-world combinatorial optimization problems at the industrial scale.

Although we have limited the scope of this work to tensor network-based generative quantum models, it would be a natural extension to consider other generative quantum models as well. For example, hybrid classical quantum models such as quantum circuit associative adversarial networks (QC-AAN) ~\cite{rudolph2020generation} can be readily explored to harness the power of generative quantum models with so-called noisy intermediate-scale quantum (NISQ) devices ~\cite{Preskill2018}. In particular, the QC-AAN framework opens up the possibility of working with a larger number of variables and going beyond discrete values (e.g., variables with continuous values).  Both quantum-inspired and hybrid quantum-classical algorithms can be tested in this GEO framework in even larger problem sizes of this NP-hard version of the portfolio optimization problem or any other combinatorial optimization problem. As the number of qubits in NISQ devices increases, it would be interesting to explore generative models that can utilize more quantum resources, such as Quantum Circuit Born Machines (QCBM)\cite{Benedetti2019}: a general framework to model arbitrary probability distributions and perform generative modeling tasks with gate-based quantum computers.

Increasing the expressive power of the quantum-inspired core of MPS to other more complex but still efficient QI approaches, such as tree-tensor networks~\cite{Cheng2019}, is another interesting research direction. Although we have fully demonstrated the relevance and scalability of our algorithm for industrial applications by increasing the performance of classical solvers on industrial scale instances (all 500 assets in the S\&P 500 market index), there is a need to explore the performance improvement that could be achieved by more complex TN representations or on other combinatorial problems.

Although the goal of GEO was to show good behavior as a general black-box algorithm without considering the specifics of the study application, it is a worthwhile avenue to exploit the specifics of the problem formulation to improve its performance and runtime. In particular, for the portfolio optimization problem with a cardinality constraint, it is useful to incorporate this constraint as a natural MPS symmetry, thereby reducing the effective search space of feasible solutions from the size of the universe to the cardinality size.

Finally, our thorough comparison with SOTA algorithms, which have been fine-tuned for decades on this specific application, shows that our TN-GEO strategy manages to outperform a couple of these and is on par with the other seven optimizers. This is a remarkable feat for this new approach and hints at the possibility of finding commercial value in these quantum-inspired strategies in large-scale real-world problems, as the instances considered in this work. 
Also, it calls for more fundamental insights towards understanding when and where it would be beneficial to use this TN-GEO framework, which relies heavily on its quantum-inspired generative ML model. For example, understanding the intrinsic bias in these models, responsible for their remarkable performance, is another important milestone on the road to practical quantum advantage with quantum devices in the near future. The latter can be asserted given the tight connection of these quantum-inspired TN models to fully quantum models deployed on quantum hardware. And this question of when to go with quantum-inspired or fully quantum models is a challenging one that we are exploring in ongoing future work.

\begin{acknowledgments} 
The authors would like to acknowledge Manuel S. Rudolph, Marta Mauri, Matthew J.S. Beach,  Yudong Cao, Luis Serrano, Jhonathan Romero-Fontalvo, and Brian Dellabetta for their feedback on an early version of this manuscript  
\end{acknowledgments}


%



\appendix

\section{Methods}
\label{s:methods}

\subsection{Generation of portfolio optimization instances}\label{ss:portfolio}

The portfolio optimization problem aims at determining the fractions $w_i$ of a given capital to be invested in each asset $i$ of a universe of $N$ assets, such that the risk $\sigma (\boldsymbol{w})$ for a given level $\rho$ of the expected return $\langle r(\boldsymbol{w}) \rangle$ is minimized, constrained to $\sum\limits_{i}^N w_i = 1$. The problem can be formulated as:
\begin{eqnarray} \label{eq:optProblem}
\min_{\boldsymbol{w}} \lbrace \sigma^2(\boldsymbol{w})   =  \boldsymbol{w^T} \cdot \boldsymbol{\Sigma} \cdot \boldsymbol{w} &:& \langle r (\boldsymbol{w}) \rangle = \boldsymbol{w} \cdot \boldsymbol{r} = \rho \rbrace
\end{eqnarray} 
where the vectors $\boldsymbol{w}$ and $\boldsymbol{r}$ have dimensionality $N$,  $\boldsymbol{\Sigma}$ is the sample covariance matrix obtained from the return time series of pair of asset $i$ and $j$, and $\boldsymbol{r}$ is the vector of average return of the time series for each asset, with each daily return, $r^t$, calculated as the relative increment in asset price from its previous day (i.e., $r^t = (p^t - p^{(t-1)}) / p^{(t-1)}$, with $p^t$ as the price for a particular asset at time $t$). The solution to Eq.~\ref{eq:optProblem} for a given return level $\rho$ corresponds to the optimal portfolio strategy $\boldsymbol{w}^*$ and  the minimal value of this objective function $\sigma(\boldsymbol{w})$ correspond to the portfolio risk and will be denoted by $\sigma^*_\rho$.

Note that the optimization task in Eq.~\ref{eq:optProblem} has the potential outcome of investing small amounts in a large number of assets as an attempt to reduce the overall risk by "over diversifying" the portfolio. This type of investment strategy can be challenging to implement in practice: portfolios composed of a large number of assets are difficult to manage and may incur in high transaction costs.  Therefore, several restrictions are usually imposed on the allocation of capital among assets, as a consequence of market rules and conditions for investment or to reflect investor profiles and preferences. For instance, constraints can be included to control the amount of desired diversification, i.e., modifying bound limits per asset $i$, denoted by $\lbrace l_i, u_i \rbrace$, to the proportion of capital invested in the investment on individual assets or a group of assets, thus the constraint $l_i < w_i < u_i$ could be considered.

Additionally, a more realistic and common scenario is to include in the optimization task a {\it cardinality constraint}, which limits directly the number of assets to be transacted to a pre-specified number $\kappa < N$. Therefore, the number of different sets to be treated is $M = \binom{N}{\kappa}$. In this scenario, the problem can be formulated as a Mixed-Integer Quadratic Program (MIQP) with the addition of binary variables $x_i \in \{0,1 \}$ per asset, for $i = 1,..., N$, which are set to ``1" when the $i$-th asset is included as part of the $\kappa$ assets, or ``0" if it is left out of this selected set. Therefore, valid portfolios would have a number $\kappa$ of $1$'s, as specified in the cardinality constraint. For example, for $N=4$ and $\kappa = 2$, the six different valid configurations can be encoded as $ \{ 0011, 0101, 0110, 1001,1010,1100 \}$.

The optimization task can then be described as follows

\begin{eqnarray} \label{eq:eqtProblemCard}
  \min_{\boldsymbol{w},\boldsymbol{x}} \left \lbrace\sigma^2(\boldsymbol{w}) \right.&& : \nonumber \\
  && \langle r(\boldsymbol{w}) \rangle = \rho , \nonumber \\   
  && l_i x_i < w_i < u_i x_i \quad i=1,...,N, \nonumber \\
  && \boldsymbol{1} \cdot \boldsymbol{x} = \left. \kappa \right\rbrace.
\end{eqnarray}
In this reformulated problem we denote by $\sigma_{\rho,\kappa}^{*}$ the minimum  portfolio risk outcome from Eq.~\ref{eq:eqtProblemCard} for a given return level $\rho$ and cardinality $\kappa$. The optimal solution vectors $\boldsymbol{w}^*$ and $\boldsymbol{x}^*$ define the portfolio investment strategy. Adding the cardinality constraint and the investment bound limits transforms a simple convex optimization problem (Eq.~\ref{eq:optProblem}) into a much harder non-convex NP-hard problem . For all the problem instance generation in this work we chose $\kappa= N/2$ and the combinatorial nature of the problems lies in the growth of the search space associated with the binary vector $\boldsymbol{x}$, which makes it intractable to exhaustively explore for a number of assets in the few hundreds. The size of the search space here is $M = \binom{N}{N/2}$

It is important to note that given a selection of which assets belong to the portfolio by instantiating $\boldsymbol{x}$ (say with a specific $\boldsymbol{x}^{(i)}$), solving the optimization problem in Eq.~\ref{eq:eqtProblemCard} to find the respective investment fractions $\boldsymbol{w}^{(i)}$ and risk value $\sigma_{\rho,N/2}^{(i)}$ can be efficiently achieved with conventional quadratic programming (QP)  solvers. In this work we used the python module cvxopt~\cite{cvxopt2020} for solving this problem.  Note that we exploit this fact to break this constrained portfolio optimization problem into a combinatorial intractable one (find best asset selection $\boldsymbol{x}$), which we aim to solve with GEO, and a tractable subroutine which can be solved efficiently with available solvers. 

The set of pairwise $(\sigma_{\rho}^{\kappa}, \rho)$, dubbed as the {\it efficient frontier}, is no longer convex neither continuous in contrast with the solution to problem in Eq.~\eqref{eq:optProblem}.

\subsection{Problem formulation for comparison with state-of-the-art algorithms}\label{apx:comparisonform}

To carry out the comparison with State-of-the-Art Algorithms, in line with the formulation used there, we generalizes the problem in Eq.~\ref{eq:eqtProblemCard} releasing the constraint of a fix level of portfolio return, instead directly incorporating the portfolio return in the objective function, encompassing now two terms: the one on the left corresponding to the portfolio risk as beforeand the one on the right corresponding to the portfolio return. The goal is to balance out both terms such that return is maximized and risk minimized. Lambda is a hyperparameter, named \textit{risk averse}, that controls if an investor wants to give more weight to risk or return. The new formulation reads as follows,

\begin{align} \label{eq:eqtProblemCardFull}
  \min_{\boldsymbol{w},\boldsymbol{x}} \lbrace\lambda\sigma^2(\boldsymbol{w}) - (1-\lambda)\langle r(\boldsymbol{w}) \rangle : \nonumber \\
  l_i x_i < w_i < u_i x_i \quad i=1,...,N, \nonumber \\
  \boldsymbol{1} \cdot \boldsymbol{x} = \left. \kappa \right\rbrace.
\end{align}

With the rest of constraints and variables definition as in Appendix \ref{ss:portfolio}.

\subsubsection{Performance Metrics}\label{apx:perfMeasure}
To compare the performance of the proposed GEO with the SOTA metaheuristic algorithms in the literature, the most commonly used performance metrics for the cardinality constrained portfolio optimization problem are used. These metric formulations compute the distance between the heuristic efficient frontier and the unconstrained efficient frontier. Thus, the performance of the algorithms can be evaluated.

Four of these performance metrics (the Mean, Median, Minimum and Maximum in Table~\ref{table:1}) are based on the so-called Performance Deviation Errors ($PDE$). These $PDE$ metrics were formulated by Chang \cite{chang2000heuristics} as follows:

\begin{equation} \label{percentage deviation error}
PDE_i = min\left(\left|\frac{100\left(x_i-x^*_i\right)}{x^*_i}\right|,\left|\frac{100\left(y_i-y^*_i\right)}{y^*_i}\right|\right)\
\end{equation}

\begin{equation} \label{eq:metric1}
    \begin{split}
    x^*_i&=X_{k_y}+\frac{\left(X_{j_y}-X_{k_y}\right)\left(y_i-Y_{k_y}\right)}{\left(Y_{j_y}-Y_{k_y}\right)} \\
    y^*_i&=Y_{k_x}+\frac{\left(Y_{j_x}-Y_{k_x}\right)\left(x_i-X_{k_x}\right)}{\left(X_{j_x}-X_{k_x}\right)} \\
    j_y&={\argmin_{l=1,\dots,{\varepsilon}^*\bigwedge Y_l\ge y_i} Y_l } \\  
    k_y&={\argmax_{l=1,\dots,{\varepsilon}^*\bigwedge Y_l\le y_i} Y_l } \\
    j_x&={\argmin_{l=1,\dots,{\varepsilon}^*\bigwedge X_l\ge x_i} X_l } \\
    k_x&={\argmax_{l=1,\dots,{\varepsilon}^*\bigwedge X_l\le x_i} X_l } \\
    \end{split}
\end{equation}

where the pair $(X_l,Y_l) (l = 1,...,\varepsilon^*)$ represents the point on the standard efficient frontier and the pair $(x_i,y_i) (i = 1,...,\varepsilon)$ represents the point on the heuristic efficient frontier. Here, $\varepsilon^*$ denotes the number of points on the standard efficient frontier while $\varepsilon$ denotes the number of points on the heuristic efficient frontier. The mean, median, minimum, and maximum of the $PDE$ can be used to compare the performance of the algorithms.

\begin{figure}
\includegraphics[width=0.48\textwidth, scale=0.25]{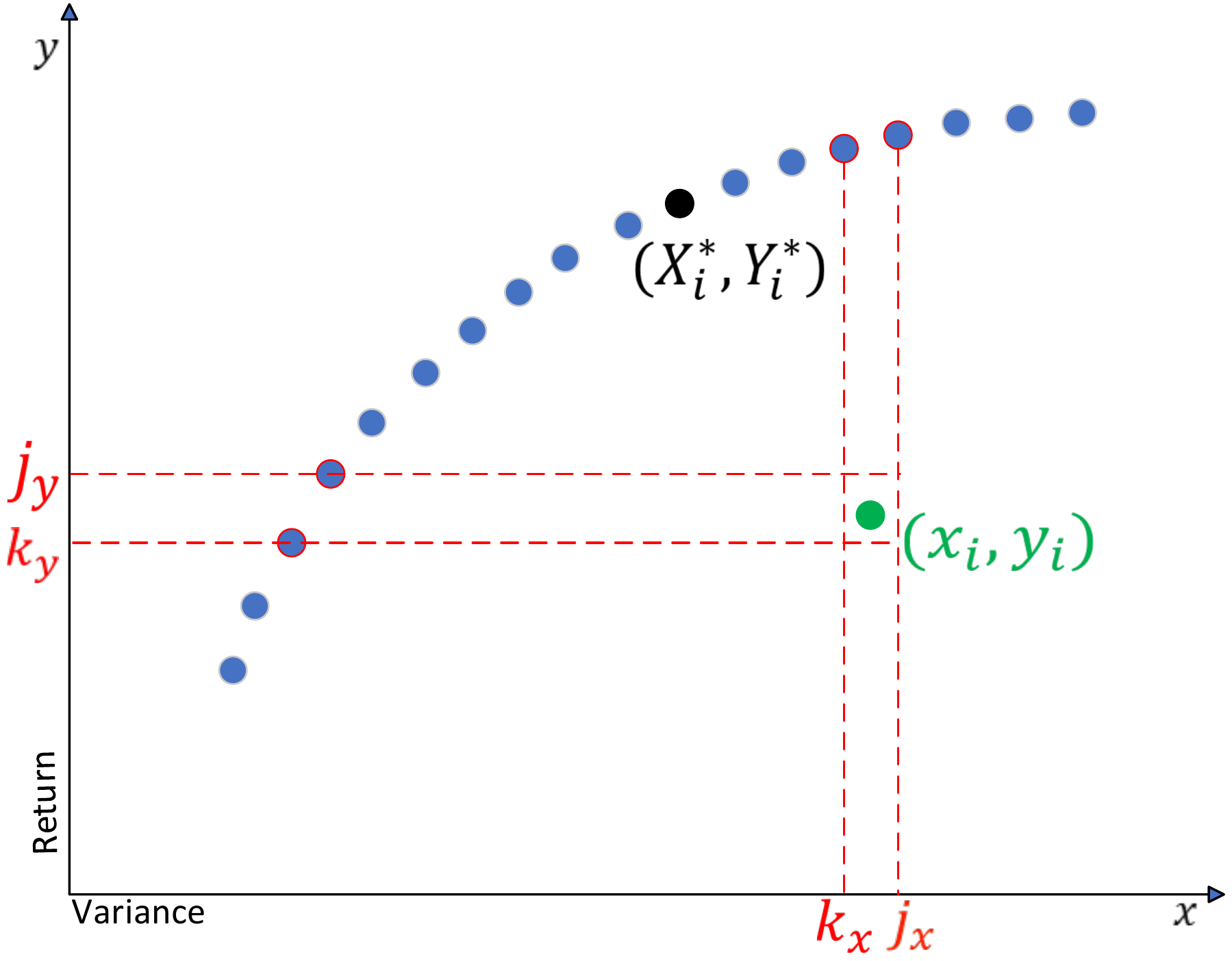}
\caption{A graphical demonstration of indices used for performance metrics calculation}
\label{fig:PerformanceMetrics}
\end{figure}

Later, three additional performance measures (MEUCD: Mean Euclidean Distance, VRE: Variance of Return Error, MRE: Mean Return Error) were formulated by Cura \cite{cura2009particle} as follows:

\begin{equation} \label{mean euclidean distance}
MEUCD=\frac{\sum^{\varepsilon}_{i=1}{\sqrt{\left(X^*_i-x_i\right)+\left(Y^*_i-y_i\right)}}}{\varepsilon }
\end{equation}
\begin{equation} \label{variance of return error}
VRE=\frac{\sum^{\varepsilon }_{i=1}{100{\left|X^*_i-x_i\right|}/{x_i}}}{\varepsilon }
\end{equation}
\begin{equation} \label{mean return error}
MRE=\frac{\sum^{\varepsilon }_{i=1}{100{\left|Y^*_i-y_i\right|}/{y_i}}}{\varepsilon }
\end{equation}
where $(X^*_i,Y^*_i)$ is the standard point closest to the heuristic point $(x_i,y_i)$. Figure \ref{fig:PerformanceMetrics} shows a graphical representation of the indices used to calculate the performance metrics for the convenience of the reader and the values for TN-GEO and all the other SOTA optimizers are reported in Table~\ref{table:1}.

\subsection{Quantum-Inspired Generative Model in TN-GEO}

The addition of a probabilistic component is inspired by the success of Bayesian Optimization (BO) techniques, which are among the most efficient solvers when the performance metric aims to find the lowest minimum possible within the least number of objective function evaluations. For example, within the family of BO solvers, GPyOpt~\cite{gpyopt2016} uses a Gaussian Process (GP) framework consisting of multivariate Gaussian distributions. This probabilistic framework aims to capture relationships among the previously observed data points (e.g., through tailored kernels), and it guides the decision of where to sample the next evaluation with the help of the so called acquisition function. GPyOpt is one of the solvers we use to benchmark the new quantum-enhanced strategies proposed here.

Although the GP framework in BO techniques is not a generative model, we explore here the powerful unsupervised machine learning framework of generative modeling in order to capture correlations from an initial set of observations and evaluations of the objective function (step 1-4 in Fig.~\ref{fig:Algo_Scheme}).

For the implementation of the quantum-inspired generative model at the core of TN-GEO we follow the procedure proposed and implemented in Ref.~\cite{Han2018}. Inspired by the probabilistic interpretation of quantum physics via Born's rule, it was proposed that one can use the Born probabilities $|\Psi(\boldsymbol{x})|^2$ over the $2^N$ states of an $N$ qubit system to represent classical target probability distributions which would be obtained otherwise with generative machine learning models. Hence,

\begin{equation}
    \label{born_rule}
    P(\boldsymbol{x}) = \frac{|\Psi(\boldsymbol{x})|^{2}}{Z} \text{, with } Z = \sum\limits_{\boldsymbol{x}\in \cal{S} }|\Psi(\boldsymbol{x})|^{2},
\end{equation}
with $\Psi(\boldsymbol{x}) = \langle \boldsymbol{x} | \Psi \rangle $ and $\boldsymbol{x} \in \{ 0,1\}^{\otimes N}$ are in one-to-one correspondence with decision variables over the investment universe with $N$ assets in our combinatorial problem of interest here. In Ref.~\cite{Han2018} these quantum-inspired generative models were named as  \textit{Born machines}, but we will refer to them hereafter as \textit{tensor-network Born machines} (TNBM) to differentiate it from the \textit{quantum circuit Born machines} (QCBM) proposal~\cite{Benedetti2019} which was developed independently to achieve the same purpose but by leveraging quantum wave functions from quantum circuits in NISQ devices. As explained in the main text, either quantum generative model can be adapted for the purpose of our GEO algorithm.

On the grounds of computational efficiency and scalability towards problem instances with large number of variables (in the order of hundreds or more), following Ref.~\cite{Han2018} we  implemented the quantum-inspired generative model based on Matrix Product States (MPS) to learn the target distributions $|\Psi(\boldsymbol{x})|^2$.

MPS is a type of TN where the tensors are arranged in a one-dimensional geometry. Despise its simple structure, MPS can efficiently represent a large number of quantum states of interest extremely well~\cite{cirac2020matrix}.  Learning with the MPS is achieved by adjusting its parameters such that the distribution obtained via Born's rule is as close as possible to the data distribution. MPS enjoys a direct sampling method that is more efficient than other Machine Learning techniques, for instance, Boltzmann machines, which require Markov chain Monte Carlo (MCMC) process for data generation.

The key idea of the method to train the MPS, following the algorithm on paper ~\cite{Han2018}, consists of adjusting the value of the tensors composing the MPS as well as the bond dimension among them, via the minimization of the negative log-likelihood function defined over the training dataset sampled from the target distribution. For more details on the implementation see Ref.~\cite{Han2018} and for the respective code see Ref.~\cite{codemps2018}.

\subsection{Classical Optimizers}\label{ss:optimizers}

\subsubsection{GPyOpt  Solver}
GPyOpt \cite{gpyopt2016} is a Python open-source library for Bayesian Optimization based on GPy and a Python framework for Gaussian process modelling. For the comparison exercise in TN-GEO as a stand-alone solver here are the hyperparameters we used for the GPyOpt solver:
\begin{itemize}
    \item Domain: to deal with the exponential growth in dimensionality, the variable space for $n$ number of assets was partitioned as the cartesian product of $n$ 1-dimensional spaces.
    \item Constraints: we added two inequalities in the number of assets in a portfolio solution to represent the cardinality condition.
    \item Number of initial data points: 10
    \item Acquisition function: Expected Improvement
\end{itemize}

\subsubsection{Simulated Annealing Solver}
For simulated annealing (SA) we implemented a modified version from Ref.~\cite{perrygeo2019}. The main change consists of adapting the update rule such that new candidates are within the valid search space with fixed cardinality. The conventional update rule of single bit flips will change the Hamming weight of $\boldsymbol{x}$ which translates in a portfolio with different cardinality. The hyperparameters used are the following:
\begin{itemize}
    \item Max temperature in thermalization: 1.0
    \item Min temperature in thermalization: 1e-4
    
\end{itemize}

\subsubsection{Conditioned Random Solver}

This solver corresponds to the simplest and most naive approach, while still using the cardinality information of the problem. In the \textit{conditioned random solver}, we generate, by construction, bitstrings which satisfy the cardinality constraint. Given the desired cardinality $\kappa = N/2$ used here, one starts from the bitstring with all zeros, $\boldsymbol{x}_{0} = 0 \cdots 0$, and flips only $N/2$ bits at random from positions containing $0$'s, resulting in a valid portfolio candidate $\boldsymbol{x}$ with cardinality  $N/2$.

\subsubsection{Random Solver}

This solver corresponds to the simplest approach without even using the cardinality information of the problem. In the \textit{random solver}, we generate, by construction, bitstrings randomly selected from the $2^N$ bitstrings of all possible portfolios, where $N$ is the number of assets in our investment universe.

\subsection{Algorithm Methodology for TN-GEO as a booster}~\label{ss:TN-GEO_booster}

As explained in the main text, in this case it is assumed that the cost of evaluating the objective function is not the major computational bottleneck, and consequently there is no practical limitations in the number of observations to be considered.

Following the algorithmic scheme in Fig.~\ref{fig:Algo_Scheme}, we describe next the details for each of the steps in our comparison benchmarks:

\begin{enumerate}
  \item \textit{Build the seed data set, \{$ \boldsymbol{x}^{(i)} \}_{\rm{seed}}$ and \{$ \sigma^{(i)}_{\rho,N/2} \}_{\rm{seed}}$}. For each problem instance defined by $\rho$ and a random subset with $N$ assets from the S\&P 500, gather all initial available data obtained from previous optimization attempts with classical solver(s). In our case, for each problem instances we collected 10,000 observations from the SA solver. These 10,000 observations corresponding to portfolio candidates \{$ \boldsymbol{x}^{(i)} \}_{\rm{init}}$ and their respective risk evaluations \{$ \sigma^{(i)}_{\rho,N/2} \}_{\rm{init}}$ were sorted and only the first $n_{\rm{seed}}=1,000$ portfolio candidates with the lowest risks were selected as the seed data set. This seed data set is the one labeled as \{$ \boldsymbol{x}^{(i)} \}_{\rm{seed}}$ and \{$ \sigma^{(i)}_{\rho,N/2} \}_{\rm{seed}}$ in the main text and hereafter. The idea of selecting a percentile of the original data is to provide the generative model inside GEO with samples which are the target samples to be generated. This percentile is a hyperparameter and we set it 10\% of the initial data for our purposes.
  \item \textit{Construct of the \textit{softmax} surrogate distribution:} Using the seed data from step 0, we construct a softmax multinomial distribution with $n_{\rm{seed}}$ classes  - one for each point on the seed data set. The probabilities outcome associated with each of these classes in the multinomial is calculated as a Boltzmann weight, $p_i = \dfrac{ e^{-\overline{\sigma}_{\rho,\kappa}^{(i)}} }{\sum\limits_{j=1}^{n_{\rm{seed}} } e^{ -\overline{\sigma}_{\rho,\kappa}^{(j)}} }$. Here, $\overline{\sigma}_{\rho,\kappa}^{(i)} = \sigma_{\rho, \kappa} ( \boldsymbol{x}^{(i)} )/T$, and $T$ is a ``temperature" hyperparameter. In our simulations, $T$ was computed as the standard deviation of the risk values of this seed data set. In Bayesian optimization methods the surrogate function tracks the landscape associated with the values of the objective function (risk values here). This \textit{softmax surrogate} constructed here by design as a multinomial distribution from the seed data observations serves the purpose of representing the objective function landscape but in probability space. That is, it will assign higher probability to portfolio candidates with lower risk values. Since we will use this softmax surrogate to generate the training data set, this bias imprints a preference in the quantum-inspired generative model to favor low-cost configurations.
  
  \item \textit{Sample from softmax surrogate}. We will refer to these samples as the training set since these will be used to train the MPS-based generative model. For our experiments here we used $n_{\rm{train}}=10000$ samples.
  
  \item \textit{Use the $n_{\rm{train}}$ samples} from the previous step to train the MPS generative model.
  
  \item \textit{Obtain $n_{\rm{MPS}}$ samples} from the generative model which correspond to the new list of potential portfolio candidates. In our experiments, $n_{\rm{MPS}} = 4000$. For the case of 500 assets, as sampling takes sensibly longer because of the problem dimension, this value was reduced to 400 to match the time in SA.
  
  \item \textit{Select new candidates:} From the $n_{\rm{MPS}}$ samples, select only those who fulfill the cardinality condition, and which have not been evaluated. These new portfolio candidates $\{ \boldsymbol{x}^{(i)} \}_{\rm{new}}$ are saved for evaluation in the next step.
  
  \item \textit{Obtain risk value for new selected samples:} Solve Eq.~\ref{eq:eqtProblemCard} to evaluate the objective function (portfolio risks) for each of the new candidates $\{ \boldsymbol{x}^{(i)} \}_{\rm{new}}$. We will denote refer to the new cost function values by  $\{ \sigma^{(i)}_{\rho,N/2} \}_{\rm{new}}$.
  
  \item \textit{Merge the new portfolios}, $\{ \boldsymbol{x}^{(i)} \}_{\rm{new}}$, and their respective cost function evaluations, $\{ \sigma^{(i)}_{\rho,N/2} \}_{\rm{new}}$ with the seed portfolios, $\{ \boldsymbol{x}^{(i)} \}_{\rm{seed}}$, and their respective cost values, $\{ \sigma^{(i)}_{\rho,N/2} \}_{\rm{seed}}$, from step 0 above. This combined super set is the \textit{new initial data set}.
  
  \item \textit{Use the new initial data set from step 7} to start the algorithm from step 1. If a desired minimum is already found or if no more computational resources are available, one can decide to terminate the algorithm here. In all of our benchmark results reported here when using TN-GEO as a booster from SA intermediate results, we only run the algorithm for this first cycle and the minima reported for the TN-GEO strategy is the lowest minimum obtained up to step 7 above.
  
 \end{enumerate}

\subsection{Algorithm Methodology for TN-GEO as a stand-alone solver}~\label{ss:TN-GEO_stand-alone}

This section presents the algorithm for the TN-GEO scheme as a stand-alone solver. In optimization problems where the objective function is inexpensive to evaluate, we can easily probe it at many points in the search for a minimum. However, if the cost function evaluation is expensive, e.g., tuning hyperparameters of a deep neural network, then it is important to minimize the number of evaluations drawn. This is the domain where optimization technique with a Bayesian flavour, where the search is being conducted based on new information gathered, are most useful, in the attempt to find the global optimum in a minimum number of steps.

The algorithmic steps for TN-GEO as a stand-alone solver follows the same logic as that of the solver as a booster described Sec.~\ref{ss:TN-GEO_booster}. The main differences between the two algorithms rely on step 0 during the construction of the \textit{initial data set} and \textit{seed data set} in step 0, the temperature use in the softmax surrogate in step 1, and a more stringent selection criteria in step 5. Since the other steps remain the same, we focus here to discuss the main changes to the algorithmic details provided in Sec.~\ref{ss:TN-GEO_booster}.

\begin{enumerate}
    \item \textit{Build the seed data set:} since evaluating the objective function could be the major bottleneck (assumed to be expensive) then we cannot rely on cost function evaluations to generate the seed data set. The strategy we adopted is to initialize the algorithm with samples of bitstrings which satisfy the hard constraints of the problem. In our specific example, we can easily generate  $n_{\rm{seed}}$ random samples, $\mathcal{D}_0 = \{ \boldsymbol{x}^{(i)} \}_{\rm{seed}}$, which satisfy the cardinality constraint. Since all the elements in this data set hold the cardinality condition, then maximum length $n_{\rm{seed}}$ of $\mathcal{D}_0$ is $\binom{N}{\kappa}$. In our experiments, we set the number of samples $n_{\rm{init}} = 2,000$, for all problems considered here up to $N=100$ assets \item \textit{Construct the softmax surrogate distribution:}  start by constructing a uniform multinomial probability distribution where each sample in $\mathcal{D}_0$ has the same probability. Therefore, for each point in the seed data set its probability is set to $p_0= 1/n_{\rm{seed}}$. As in TN-GEO as a booster, we will attempt to generate a softmax-like surrogate which favors samples with low cost value, but we will slowly build that information as new samples are evaluated. In this first iteration of the algorithm, we start by randomly selecting a point  $\boldsymbol{x}^{(1)}$ from  $\mathcal{D}_0$, and we evaluate the value of its objective function $\sigma^{(1)}$ (its risk value in our specific finance example). To make this point $\boldsymbol{x}^{(1)}$ stand out from the other unevaluated samples, we set its  probability to be twice that of any of the remaining $n_{\rm{seed}} - 1$ points in $\mathcal{D}_0$. Since we increase the probability of one of the points, we need to adjust the probability of the $n_{\rm{seed}} - 1$ from $p_0$ to $p^{'}_0$, and if we assume the probability weights for observing each point follows a multinomial distribution  with Boltzmann weights, under these assumptions, and making by fixing the temperature hyperparameter we can solve for the reference ``risk" value $\sigma^{(0)}$ associated to all the other $n_{\rm{seed}} - 1$ points as shown below. It is important to note that $\sigma^{(0)}$ is an artificial reference value which is calculated analytically and does not require a call to the objective function (in contrast to $\sigma^{(1)}$). Here, $\mathcal{N}$ is the normalization factor of the multinomial and $T$ is the temperature hyperparameter which, as in the case of TN-GEO as a booster, can be adjusted later in the algorithm as more data is seen. Due to the lack of initial cost function values, in order to set a relevant typical ``energy" scale in this problem, we follow the procedure in Ref.~\cite{Alcazar2020ClvsQuant} where it is set to be the square root of the mean of the covariance matrix defined in Eq.~\ref{eq:optProblem}, as this matrix encapsulates the risk information (volatility) as stated in the Markowitz's model.

\begin{equation} \label{eq:Eq_Prior_MPS_Step2}
   	\begin{array}{l}
	\begin{cases}
		(n_{\rm{seed}}-1) p_{0}^{'}  + p_{1}= 1 \\
		p_{1} = 2 \cdot p_{0}^{'}
    \end{cases}
	\Rightarrow 
	\begin{cases}
		p_{0}^{'} = 1 / (1 + n_{\rm{seed}})  \\
		p_{1} = 2 / (1 + n_{\rm{seed}})
    \end{cases}
    \\
    \\
	\begin{cases}
		\mathcal{N} =  (n_{\rm{seed}}-1) e^{-\sigma^{(0)}/T} + e^{-\sigma^{(1)}/T}
		\\
		p_{1} = e^{-\sigma^{(1)}/T}/\mathcal{N}\\
		p_{0}^{'} = e^{-\sigma^{(0)}/T}/\mathcal{N}\\
    \end{cases}
	\Rightarrow
	\\
	\\
	\begin{cases}
		\mathcal{N} =  (n_{\rm{seed}}+1)\cdot e^{-\sigma^{(1)}/T} / 2
		\\
		\sigma^{(0)} = T\cdot \log{2} + \sigma^{(1)}\\
    \end{cases}
	\end{array}
  \end{equation}
  
  \item \textit{Generate training set:} same as in TN-GEO as a booster (see Appendix~\ref{ss:TN-GEO_booster}).

    \item \textit{Train MPS:} same as in TN-GEO as a booster (see Appendix~\ref{ss:TN-GEO_booster}).
    
    \item \textit{Generate samples from trained MPS:} same as in TN-GEO as a booster (see Appendix~\ref{ss:TN-GEO_booster}).
    
    \item \textit{Select new candidates from trained MPS:} In contrast to TN-GEO as a booster we cannot afford to evaluate all new candidates coming from the MPS samples. In our procedure we selected only two new candidates which must meet the cardinality constraint. For our procedure these two candidates correspond to the most frequent sample (``exploitation") and the least frequent sample (``exploration"). If all new samples appeared with the same frequency, then we can select two samples at random. In the case where no new samples were generated, we choose them from the unevaluated samples of the original seed data set in $\mathcal{D}_{0}$
    
    \item \textit{Obtain risk value for new selected samples:}  same as in TN-GEO as a booster (see Appendix~\ref{ss:TN-GEO_booster}).

   \item \textit{Merge the new portfolios with seed data set from step 0}  same as in TN-GEO as a booster (see Appendix~\ref{ss:TN-GEO_booster}).

   \item \textit{Restart next cycle of the algorithm with the merge data set as the new seed data set:}  same as in TN-GEO as a booster (see Appendix~\ref{ss:TN-GEO_booster}).

\end{enumerate}


\section{Relative TN-GEO Enhancement}\label{apx:relQE}
\begin{figure*}[h!]
\includegraphics[width=0.8 \textwidth, scale=1.0]{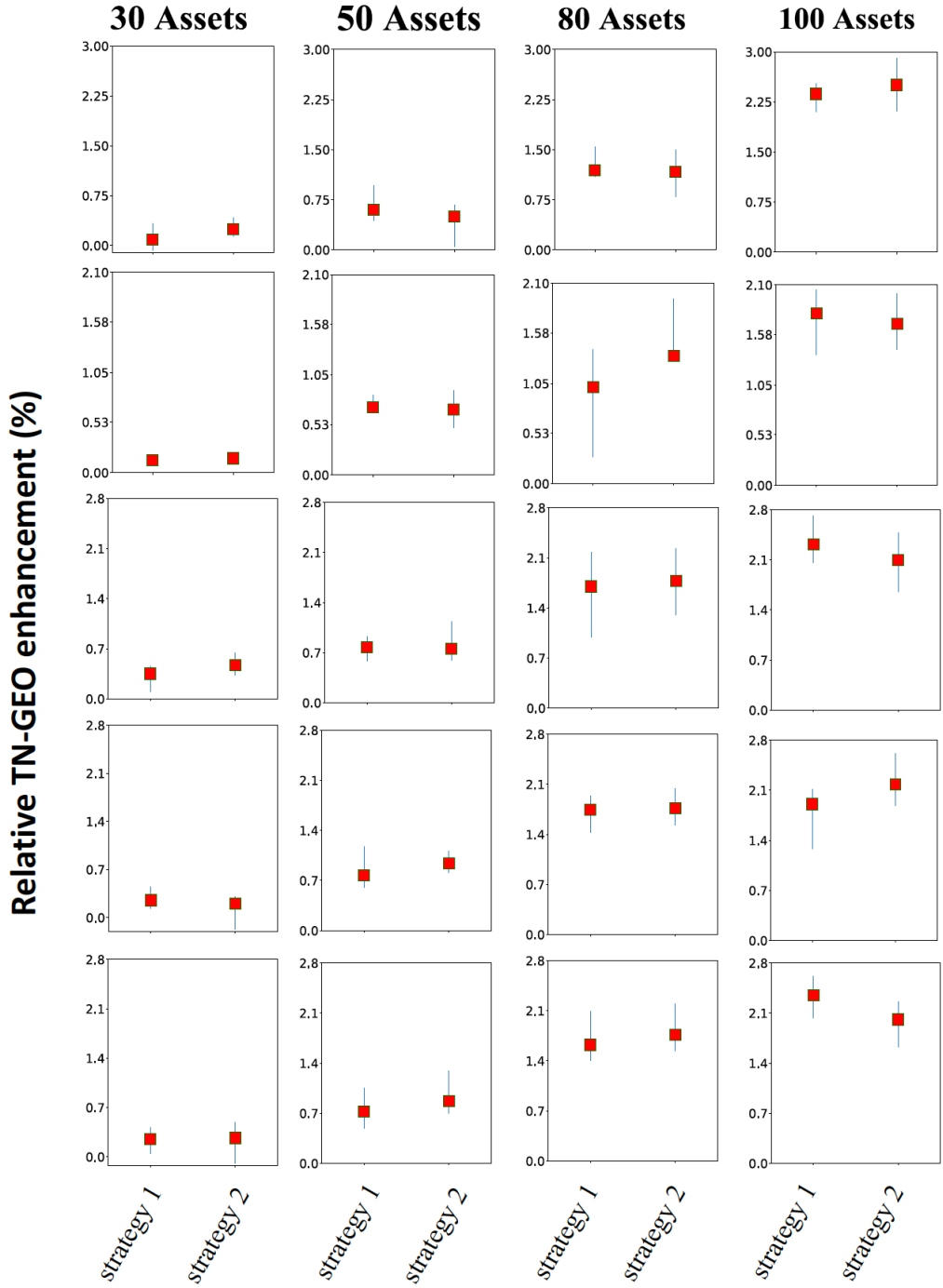}
\caption{Relative TN-GEO enhancement similar to those shown in the bottom panel of  Fig.~\ref{fig:TN-GEObooster} in the main text. For these experiments, portfolio optimization instances with a number of variables ranging from $N=30$ to $N=100$ were used. Here, each panel correspond to a different investment universes corresponding to a random subset of the S\&P 500 market index. Note the trend for a larger quantum-inspired enhancement as the number of variables (assets) becomes larger, with the largest enhancement obtained in the case on instances with all the assets from the S\&P 500 ($N=500$), as shown in Fig.~\ref{fig:TN-GEObooster}}
\label{fig:Solver2_AllResults_AllSeeds}
\end{figure*}

Figure \ref{fig:Solver2_AllResults_AllSeeds} represents the relative performance within the strategies 1 and 2 referred to subsection \ref{subsectionTNQEO}.

\clearpage

\end{document}